\documentclass[%
 amsmath,amssymb,
 aps, prl, reprint,
]{revtex4-1}

\usepackage{graphicx}
\usepackage{color}
\usepackage{dcolumn}
\usepackage{bm}
\usepackage[table]{xcolor}
\usepackage{booktabs}
\usepackage{colortbl,array} 
\usepackage{multirow,bigdelim}

\begin{document}

\title{Free energy, friction, and mass profiles from short molecular dynamics trajectories}

\author{Andrea P\'erez-Villa}
\affiliation{Sorbonne Universit\'e, Mus\'eum National d'Histoire Naturelle, UMR CNRS 7590, IRD, Institut de Min\'eralogie, de Physique des Mat\'eriaux et de Cosmochimie, IMPMC, F-75005 Paris, France}

\author{Fabio Pietrucci}
\email{fabio.pietrucci@sorbonne-universite.fr}
\affiliation{Sorbonne Universit\'e, Mus\'eum National d'Histoire Naturelle, UMR CNRS 7590, IRD, Institut de Min\'eralogie, de Physique des Mat\'eriaux et de Cosmochimie, IMPMC, F-75005 Paris, France}

\date{\today}

\begin{abstract}
We address the problem of constructing accurate mathematical models of the dynamics of 
molecular systems projected on a collective variable.
To this aim we introduce an algorithm optimizing the parameters of a standard or generalized Langevin equation until the latter reproduces in a faithful way a set of molecular dynamics trajectories.
In particular, using solvated proline dipeptide as a test case,
we report evidence that $\sim$100 short trajectories 
initiated at the top of a high barrier encode all the information needed to reconstruct 
free energy, friction, and mass profiles, including non-Markovian effects.
The approach allows accessing the thermodynamics and kinetics of activated processes
in a conceptually direct way, it employs only standard unbiased molecular dynamics trajectories, and is competitive in computational cost with respect to existing enhanced sampling methods.
Furthermore, the systematic construction of Langevin models for different choices of collective variables starting from the same initial data could help in reaction coordinate optimization. 
\end{abstract}

\pacs{Valid PACS appear here}
\maketitle



The accurate characterization of rare events like phase transitions, chemical reactions, or biomolecular conformational changes is one of the primary aims of computer simulation. The corresponding free energy barriers can often be reconstructed by means of the many available enhanced sampling techniques based on molecular dynamics or Monte Carlo simulations \cite{Pietrucci17review}. The latter calculations are computationally expensive; however, the quantitative prediction of kinetic properties like transition rates is even more difficult, and it often rests on approximations -- sometimes inadequate -- like those of transition state theory, classical nucleation theory, Markovian dynamics, etc. Algorithms addressing the kinetics or rare events are comparatively less abundant and developed than those for free-energy calculations, and they typically require the production of very extensive simulation data sets \cite{Camilloni18}. 

From a theoretical viewpoint, the framework of free-energy landscapes with barriers separating metastable states, customarily invoked to interpret rare events, corresponds to the analysis of equilibrium probability distributions as a function of a small number of collective variables (CVs). This framework, in turn, is contained in the framework of Langevin equations \cite{Zwanzig01,Risken96}, able to approximate the equilibrium and out-of-equilibrium (e.g., relaxation to equilibrium) dynamics of the high-dimensional many-particle system projected on CVs, yielding equilibrium probability distributions as a by-product. When employing the optimal reaction coordinate as CV, Langevin equations provide also the exact mean first passage times (inverse of transition rates) and mean transition path times:
such optimal coordinate corresponds to the committor function, associating to any configuration the probability to evolve towards products before reaching reactants \cite{Banushkina16,Peters16}.

The high-dimensional dynamics of many-particle systems projected on a CV $x$ can be modeled by a non-Markovian, generalized Langevin equation (GLE): \cite{Zwanzig01,Luczka05}
\begin{equation}
m \ddot x = -\frac{dF}{dx} -m \int_0^{\infty} dt'\ \Gamma(t')\ \dot x (t-t') + R(t)
\end{equation}
where $m$ is the mass, $F$ is the free-energy profile (the potential of mean force), $\Gamma(t)$ is the memory kernel of the friction force: all three can be position dependent. $R(t)$ is a random force with zero mean and with correlation given by the fluctuation-dissipation theorem: $\left< R(0) R(t) \right> = mk_BT\cdot \Gamma(t)$.
In systems like small solutes immersed in a bath of liquid molecules, memory effects in the friction and noise are necessary to reproduce the correct dynamics \cite{Grote80,Lee15,Daldrop18},
whereas in other applications such effects are often neglected obtaining the memory-less standard Langevin equation (SLE):
\begin{equation}
m\ddot x = -\frac{dF}{dx} -m\gamma\dot x + R(t)
\end{equation}
where now $\left< R(0) R(t) \right> = mk_BT\cdot 2 \gamma \delta(t)$ and $\gamma=\int_0^{\infty} dt\ \Gamma(t)$ is the friction coefficient.
Furthermore, when friction is very large and the velocity $\dot x$ conforms to the equilibrium distribution at all observed times, the overdamped form of the Langevin equation is an appropriate approximation. 

Several algorithms aim at constructing an optimal Langevin equation starting from dynamical trajectories of many-particle systems \cite{Straub87,Timmer00,Gradivsek00,Chorin02,Hummer03,Best06,Lange06,Horenko07,Darve09,Micheletti08,Lee15,Schaudinnus15,schaudinnus16,Lesnicki16,Meloni16,Daldrop18}. Customarily, the profiles entering the Langevin equation are estimated as equilibrium ensemble averages $\left< ... \right>$ of different functions. For instance, $F(x)$ can be computed from the histogram of the position $-k_BT \log P(x)$, and $\Gamma(t)$ from correlation functions of velocity, acceleration, and mean force \cite{Daldrop18}: ergodic sampling in brute force MD simulations can however be attained only for barriers limited to a few $k_BT$, strongly limiting the scope of this kind of techniques.  
For this reason, so far Langevin equations have been widely employed mostly as benchmark and illustrative models, instead of a routine tool to accurately reconstruct the dynamics of many activated processes in condensed matter. 

In this work we present a conceptually simple and computationally efficient method to construct optimal Langevin models of rare events in complex systems, irrespective of the height of the barrier separating metastable states.
The models provide accurate thermodynamic and kinetic information, including free energy profiles and transition rates, about the original systems.
Our main result is that a limited number ($100-1,000$) of short, unbiased trajectories relaxing from the top of a barrier, projected on a suitable CV, encode all the necessary information to accurately reconstruct the free energy profile, the friction profile and the mass profile.

Our optimization strategy has some analogies with the ones of Ref. \cite{Biswas18}, based on sets of trajectories shooted from configurations explored with metadynamics \cite{Laio02}, and of Ref. \cite{Innerbichler18}, that addressed water nucleation based on the Bayesian technique of Ref. \cite{Hummer05}: in our case, however, we directly compare MD trajectories with Langevin trajectories, without resorting to a discrete master equation, thus avoiding the space-time discretization errors inherent in the construction of Markov state models. For the same reason, our technique is not limited to the overdamped regime but it naturally encompasses SLE as well as non-Markovian GLE, i.e., the natural outcome of projecting many-body dynamics on a single CV \cite{Zwanzig01}.

Our approach requires a preliminary identification of transition state configurations along reactive pathways, using one of the many effective techniques available to this task \cite{Izrailev99,Bolhuis02,Laio02,Best05,Samanta14}: although not trivial, this step is generally much less involved than reconstructing free energy and friction profiles using the available methods. 
The algorithm we propose is straightforward:
\begin{enumerate}
\item Starting from a configuration committed to reactants and products with $\approx50$\% probability, a set of short MD trajectories relaxing to the free energy wells is generated, a CV $x$ is chosen and the probability distribution $P_\mathrm{MD}(x,v,t)$ is estimated as a normalized histogram, with $v \equiv \dot x$.
\item A set of Langevin trajectories $x(t)$ of the same duration is generated from given $F(x)$, $\Gamma(x,t)$, $m(x)$ profiles, and $P_\mathrm{LE}(x,v,t)$ is estimated as a normalized histogram. 
\item The profiles in the Langevin equation are systematically varied, minimizing the following deviation of the model with respect to the reference data:
\end{enumerate}
\begin{equation}
\epsilon=\int dx\ dv\ dt\ [P_\mathrm{LE}(x,v,t)-P_\mathrm{MD}(x,v,t)]^2
\label{eq:error}
\end{equation}
For both MD and Langevin trajectories, the initial conditions correspond to $x(t=0)=x_\mathrm{TS}$ (i.e., the transition state value) and to $v(t=0)$ randomly drawn from the canonical distribution:  $P_\mathrm{MD}(x,v,0)=P_\mathrm{LE}(x,v,0)=C\delta(x-x_\mathrm{TS})e^{-mv^2/2k_BT}$ ($P_\mathrm{LE}(x,v,t)$ is the solution of the Fokker-Planck equation associated to the Langevin equation).
After error minimization, the Langevin model reproduces in an optimal way the original MD data, yielding an estimate of the free energy, friction, and mass profiles as a function of the CV.
The error $\epsilon$ can be minimized employing the following simple Monte Carlo algorithm (or a more sophisticated scheme): $F(x)$, $\Gamma(x,t)$, and $m(x)$ profiles are parametrized by cubic splines (typically with 5 to 9 control points), and random moves are performed in parameter-space with an acceptance probability equal to $\min (1,e^{-(\epsilon_{\mathrm{new}}-\epsilon_{\mathrm{old}})/\delta})$, progressively reducing $\delta$ (see SI for details). Due to the stochastic nature of the data, in all the following applications the final estimate is taken as the average over ten independent optimizations.

\begin{figure}
\centering
\includegraphics[width=7cm]{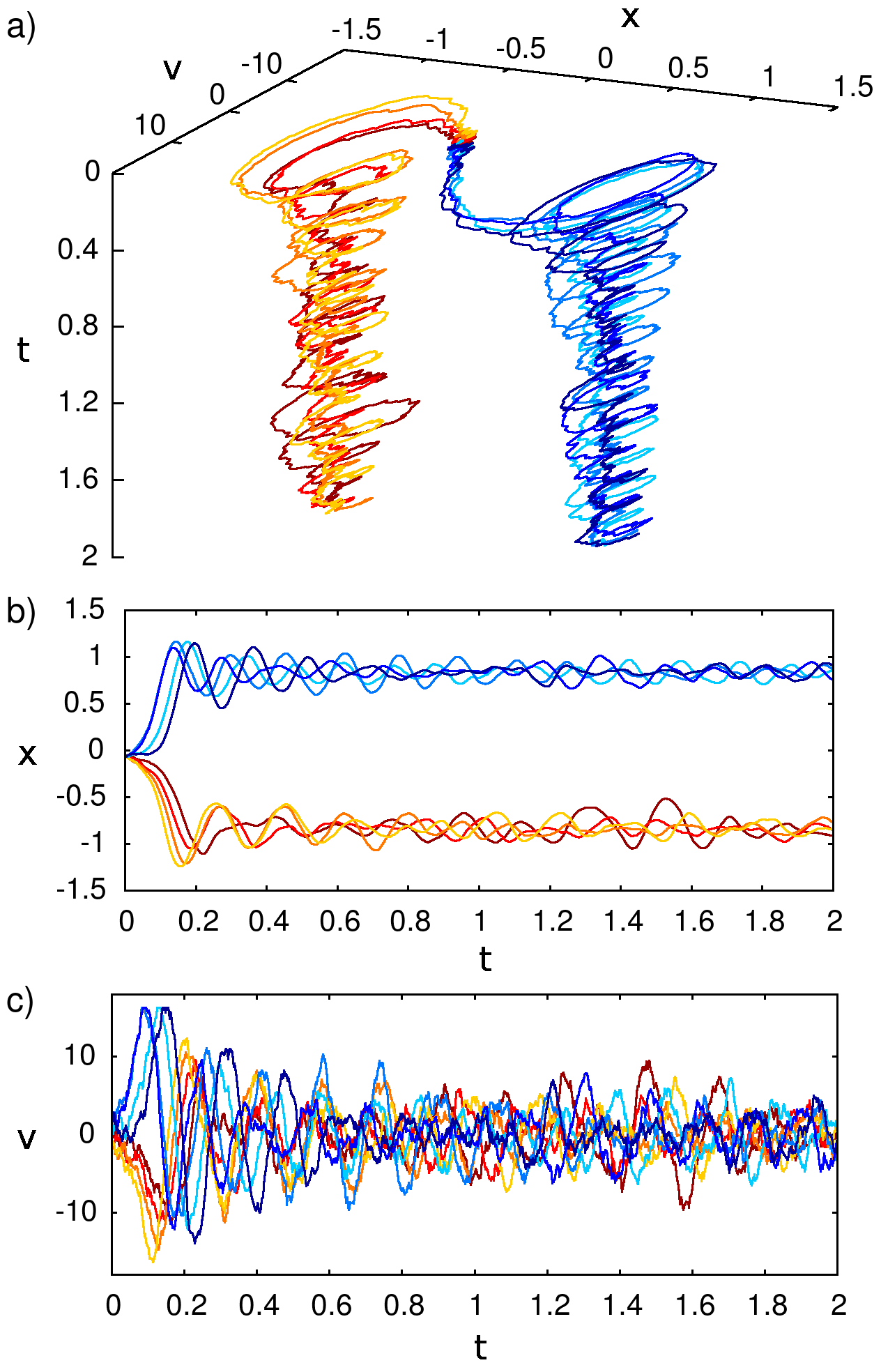}
\caption{
Examples of trajectories relaxing from the top of a barrier towards either the left well (red lines, depth 20 $k_BT$) or right well (blue lines, depth 40 $k_BT$) in the free energy landscape of Fig. \ref{fig2}-b. Such kind of trajectories are the sole input (together with the temperature) of the novel approach described in this work. As shown in panel a), the trajectories initially accelerate, then they spiral towards the center of the wells while decelerating. Panels b) and c) display projections on the $(x,t)$ and $(v,t)$ planes, respectively.
}\label{fig1}
\end{figure}

Clearly, a decent initial guess of the parameters can facilitate the optimization: we adopt the following strategy, that does not require any additional information besides the projected MD trajectories $x(t)$ and the temperature. 
We initialize the mass profile at the constant value $m=k_BT/\left<v(t=0)^2\right>$, based on energy equipartition and on the fact that initial atomic velocities, by construction, have a canonical distribution.
For the initial friction profile $\gamma$ we adopt a constant value equal to the average of $\mathrm{var}(x)/\tau_\mathrm{corr}$ ($\tau_\mathrm{corr}$ being the autocorrelation time of $x$) estimated separately in the two wells using the last, oscillating part of the trajectories. Note that the latter formula is obtained in Ref. \cite{Hummer05} for the overdamped case, but is here employed as an approximate guess also for SLE and GLE. We adopt here the simple form 
$\Gamma(t) = \frac{\gamma}{\tau}e^{-t/\tau}$ for the memory kernel, that proved a good approximation in the case of the dihedral dynamics of solvated dialanine \cite{Lee15} and butane \cite{Daldrop18}.
$\tau$ can be arbitrarily initialized to a value of the order of the timestep. Alternatively, it is also possible to estimate initial $\gamma$ and $\tau$ values by applying the more complex algorithm of Ref. \cite{Daldrop18} to trajectories equilibrated in the two wells.
Finally, we initialize the free energy using a set of ten arbitrary double-well profiles with barrier height between 10 and 100 $k_BT$.
For all systems simulated, we adopted the same time step $\delta t=0.002$ ps for MD and Langevin trajectories as well as for the estimation of the probability histograms, and we generate \cite{Fox88} a large number of Langevin trajectories ($10^4 - 10^6$) to reconstruct precisely $P_\mathrm{LE}(x,v,t)$, thus reducing the noise in the calculation of $\epsilon$ (see SI for details).

\begin{figure}
\centering
\includegraphics[width=7cm]{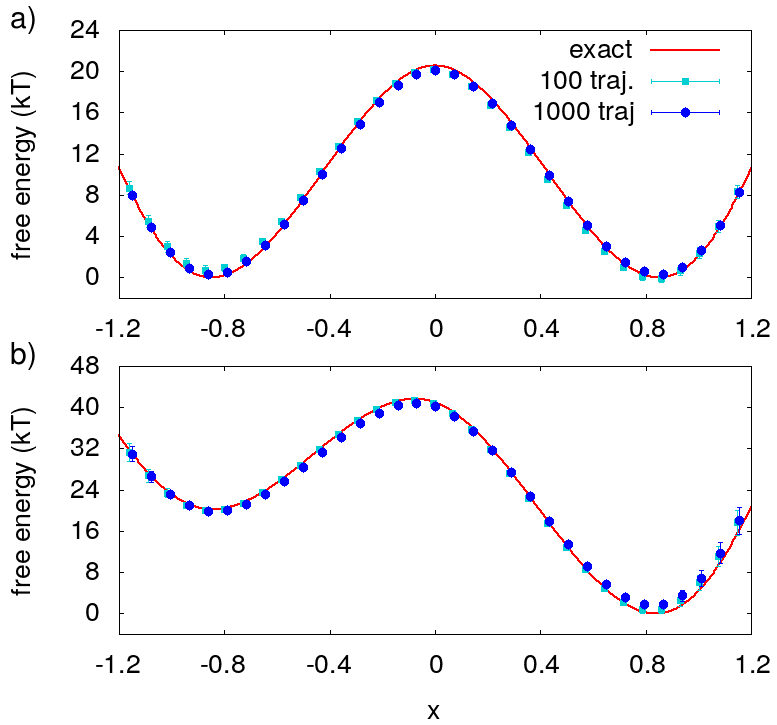}
\caption{
Optimal free energy profiles reconstructed from 100 (cyan squares) or 1,000 (blue circles) reference SLE trajectories of 2 ps, relaxing from the barrier top of a) symmetric or b) asymmetric double-wells. 
Error bars correspond to standard deviations over 10 independent optimization runs.
The exact profiles are depicted with red lines.
The exact values of the (position-independent) friction and mass are $\gamma=10$ ps$^{-1}$, $m=0.1$ kcal/mol ps$^2$;
the reconstructed values are
a) $\gamma=10.0$ and $m=0.090$ (100 traj.), 
   $\gamma=10.3$ and $m=0.096$ (1,000 traj.), 
b) $\gamma=10.2$ and $m=0.099$ (100 traj.),
   $\gamma=10.6$ and $m=0.097$ (1,000 traj.).
}\label{fig2}
\end{figure}

As a first benchmark we apply the new technique to reconstruct the free energy profile, friction and mass corresponding to a double-well SLE model with position independent $\gamma=10$ or 100 ps$^{-1}$ and $m=0.1$ kcal/mol ps$^2$ ($x$ is here adimensional). We consider barriers of 20 or 40 $k_BT$, at $T=300$ K, including both symmetric and asymmetric double wells (described by five spline control points).
In all cases, the only input information are 2 ps-long trajectories $x(t)$ of the CV relaxing from the barrier top towards the two wells (committor $= 50\pm10\%$), used as reference data for the optimization: see Fig. \ref{fig1} for examples of the evolution of $x$ and $v$.
The optimization procedure converges smoothly and rapidly ($<10^5$ Monte Carlo steps) to an excellent approximation of the exact results: as few as 100 reference trajectories are sufficient to reconstruct the free energy profile to within 1 $k_BT$ and $\gamma$, $m$ to within 10\% error. Increasing to 1,000 reference trajectories yields minimal improvements (see Fig. \ref{fig2}). The convergence behavior is similar in the small and large friction cases (see SI for details).

\begin{figure}
\centering
\includegraphics[width=7cm]{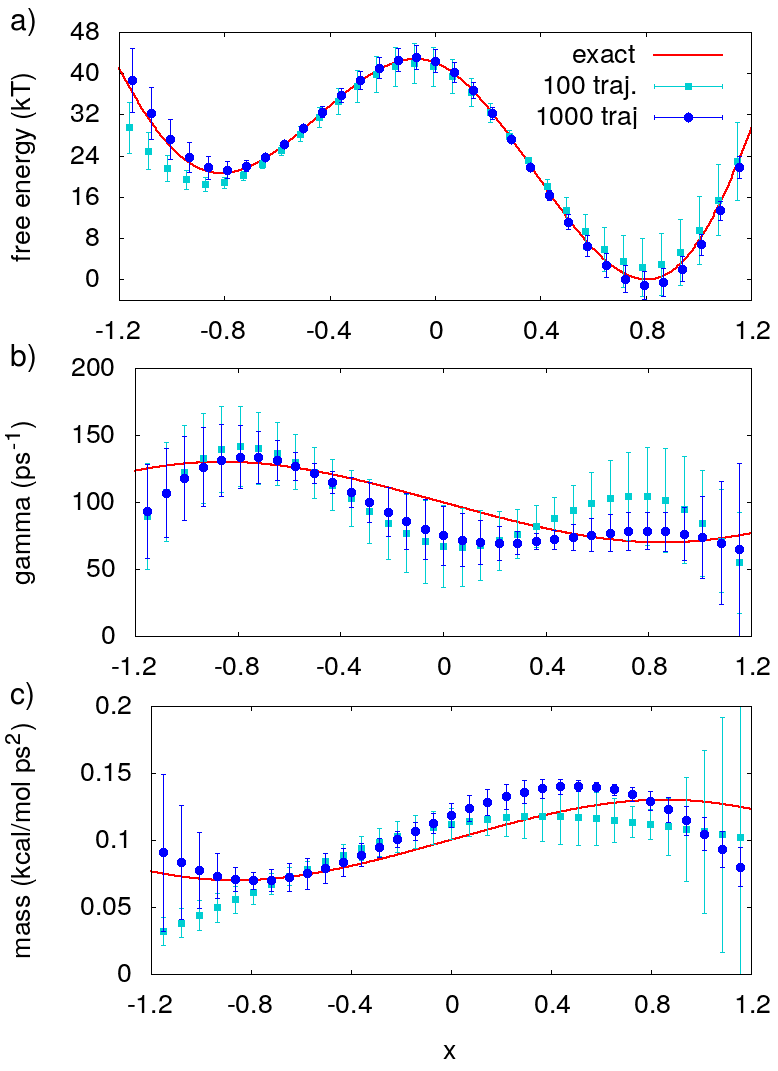}
\caption{
Optimal a) free energy, b) friction, and c) mass profiles reconstructed based on 100 (cyan squares) or 1,000 (blue circles) reference GLE trajectories of 2 ps, relaxing from the barrier top. 
Error bars correspond to standard deviations over 10 independent optimization runs.
The exact profiles are depicted with red lines.
The exact time constant of the memory kernel is $\tau=0.07$ ps, while the reconstructed values are $0.063 \pm 0.009$ ps and $0.067 \pm 0.004$ ps based on 100 or 1,000 trajectories, respectively.
}\label{fig3}
\end{figure}

Next, as a considerably more difficult benchmark, we analyze reference trajectories generated with a non-Markovian GLE, featuring  position-dependent friction and mass profiles, as shown in Fig. \ref{fig3}.
$\gamma(x)$ varies between 70 and 130 ps$^{-1}$, $m(x)$ between 0.07 and 0.13 kcal/mol ps$^2$, and the friction time constant is $\tau=0.07$ ps.
The higher complexity of the system, due to memory effects and to a considerable number of parameters to optimize (5 spline control points for each of the three profiles, plus $\tau$), renders more arduous the convergence of the optimization process: 
nevertheless, a clear correlation is observed between reduction of the error $\epsilon$ and reduction of the difference between exact and estimated parameters. Within 10$^6$ Monte Carlo steps the optimal model reaches the good accuracy shown in Fig. \ref{fig3}. 
As in previous benchmarks, employing 1,000 reference trajectories reduces the error, for a same number of optimization steps, compared to the case of 100 trajectories, but the improvement is not dramatic.

\begin{figure}
\centering
\includegraphics[width=7cm]{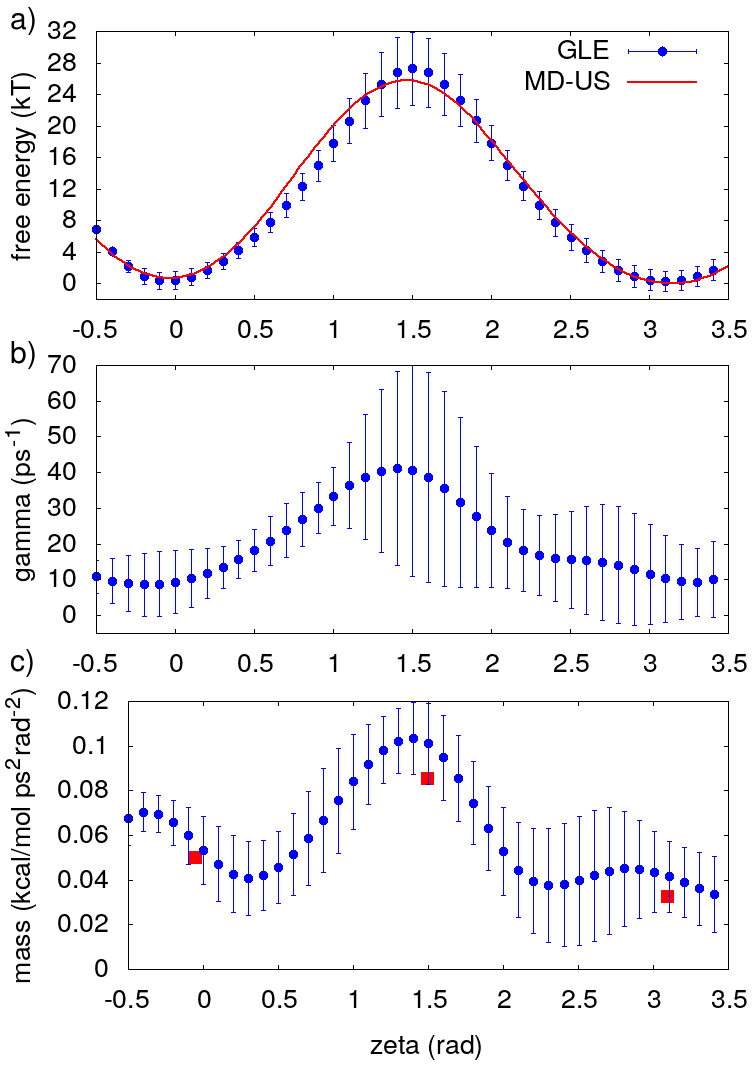}
\caption{
Optimal a) free energy, b) friction, and c) mass profiles reconstructed from 100 reference 3 ps-long MD trajectories of proline dipeptide in solution, relaxing from the barrier top.
Error bars correspond to standard deviations over 10 independent optimization runs.
Red indicates a) the free energy profile computed with umbrella sampling, and c) the mass estimated from energy equipartition (see text).
The memory kernel time constant is estimates as $\tau=0.017 \pm 0.014$ ps.
}\label{fig4}
\end{figure}

Finally, as a realistic application to a condensed matter system, we analyzed the cis/trans isomerization of a proline dipeptide Ace-Pro-Nme (AMBER03 force field \cite{Ponder03}) solvated with 502 TIP3P~\cite{Jorgensen1983} water molecules. Isomerization of prolyl peptide bonds is a crucial process: a rate-limiting step in folding, it affects protein stability, denaturation, epigenetic modifications, and so forth~\cite{Wedemeyer2002,Taylor2003}. All-atom molecular dynamics simulations were performed with a timestep of 0.002 ps at 300 K in the canonical ensemble (stochastic velocity rescaling~\cite{Bussi2007} with time constant = 0.1 ps).
As CV we consider the zeta dihedral angle (CH$_3$-O$_1$-C$_\delta$-C$_R$) in the range $[-0.5,3.5]$ rad, and to establish a reference free energy profile we performed both well-tempered metadynamics \cite{Barducci08} and umbrella sampling simulations \cite{Roux95}, limiting the psi dihedral angle (N$_1$-C$_R$-C-N) between $[-2.094, 0.785]$ rad to prevent transitions orthogonal to zeta (see SI for all technical details). The two techniques provide the same profile within 1 $k_BT$, featuring an almost symmetric double well with a barrier of 26 $k_BT$ (Fig. \ref{fig4}), in good agreement with Ref.~\cite{Melis2009}.

The zeta coordinate involves only 4 atoms and is not expected to be an optimal reaction coordinate, due to the exclusion of remaining peptide atoms and of solvent degrees of freedom. This is a typical situation in enhanced sampling simulations, hence a good testing ground to assess the performance of our optimization algorithm when projecting high-dimensional MD data on a CV. To this aim, we extracted from the metadynamics trajectory a set of configurations along the first crossing of the cis/trans isomerization barrier, and identified by committor analysis a configuration relaxing 53 times to the left well and 47 to the right (zeta=1.491, psi=-0.544).
The latter set of 100 relaxation trajectories of 3 ps is used as reference to optimize a Langevin model.
Analysis of the velocity probability distribution on the sub-picosecond time scale (see SI) indicates strong deviations from the canonical distribution, ruling out overdamped dynamics. 
We optimize a GLE model, able to account for inertial and memory effects on the dynamics, consistently with recent studies of small-molecule isomerization in solution \cite{Lee15,Daldrop18}, describing each profile with 9 spline control points.

As shown in Fig. \ref{fig4}, the optimized free energy profile reproduces within less than 2 $k_BT$ the umbrella sampling reference. Both the friction and mass profiles display a sizable modulation with the position, that cannot be discarded during optimization without spoiling the accuracy of the free energy profile, and in particular the mass profile is consistent with the values estimated from $m=k_BT/\left< v^2 \right>$ at the barrier top (using $v(t=0)$ from MD relaxation trajectories) and in the two minima (using equilibrated MD trajectories of 200 ps). 
The optimal model yields $\tau \approx 17$ fs for the exponential memory kernel, so that non-Markovian effects cannot be neglected on the fine resolution of 2 fs adopted to integrate and analyze the Langevin equation.

Two important issues remain to be understood. 
First, when projecting high-dimensional MD trajectories on one CV, a number of different atomic configurations with transition state-like behavior can be employed as starting point for the relaxation trajectories: what is the effect of such variability on the reconstructed Langevin models? To start addressing this question we generated 100 trajectories from a second transition state configuration at zeta=1.428, psi=-0.620, obtained performing a 100 ps umbrella sampling simulation centered on the first transition state, with bias $=320\,(zeta-1.49)^2+160\,(psi+0.544)^2$ $k_BT$, followed by committor analysis.
Optimization of a new GLE model (see SI) leads once again to a free energy profile and mass profile consistent with the reference results within statistical error bars. However, the latter are quite large, especially in the case of the friction profile: work in progress is devoted to enhancing the Monte Carlo minimization scheme.

A second issue concerns the effect of using sub-optimal CVs, rather than the ideal reaction coordinate, for Langevin model optimizations. Future work will assess if the CV definition is amenable to optimization through an efficient scheme, where a single initial set of MD relaxation trajectories is employed to build many different optimal Langevin models for different CVs, the definition of the latter being iteratively improved based on the analysis of the model's kinetic properties. A related idea was recently proposed in the case of discrete Markov state models \cite{Tiwary16}. 


To summarize, the evidence presented in this work points to the non-trivial conclusion that $100 - 1,000$ short MD trajectories relaxing from the top of a high barrier encode all the information necessary to reconstruct complete free energy, friction and mass profiles, both in the Markovian and non-Markovian cases. Remarkably, all this is achieved with recourse neither to long ergodic trajectories nor to external biasing forces, but employing short trajectory segments naturally drifting towards low free-energy regions. Such trajectories consist in time evolution unhampered by barriers, hence the MD computational cost is limited by the intrinsic transition path time and appears close to the theoretical minimum from an intuitive enhanced-sampling viewpoint.

It is important to note that the reactive flux formalism \cite{Chandler78}, providing transition rates with a correct transmission coefficient, can be seamlessly combined with the present approach without extra computational cost, since the required correlation function can be estimated using inexpensive Langevin trajectories. This leads to the characterization of the system's kinetics on arbitrarily long time scales, one of the most desirable high-hanging fruits of atomistic simulations.

Encompassing Langevin equations ranging from the overdamped to the inertial to the non-Markovian memory-friction regime, the new method could be potentially applied to a very wide range of activated processes, from ice nucleation to biomolecular conformational changes to chemical reactions in solution. Clearly, improved algorithms for the exploration of transition state ensembles would be beneficial in combination with the present method.
Finally, the fact that a single set of reference MD data can be exploited to construct Langevin models in a systematic way for different choices of CV might also facilitate the application and development of reaction coordinate optimization techniques.

\subsection{Acknowledgements}

We gratefully acknowledge very insightful discussions with A. Marco Saitta, Rodolphe Vuilleumier, Alessandro Laio, Riccardo Ferrando and Gerhard Stock.


\bibliography{biblio}
\end{document}